\documentclass[aip,prb,showpacs,twocolumn]{revtex4}

\usepackage{graphics,graphicx,amssymb,amsmath,color}
\begin{document}

\title{Refining the Spin Hamiltonian in the Spin-$\frac{1}{2}$ Kagome Lattice Antiferromagnet ZnCu$_{3}$(OH)$_{6}$Cl$_{2}$ using Single Crystals}
\author{Tianheng~Han$^{1,\ddag}$}
\author{Shaoyan~Chu$^{2}$}
\author{Young S.~Lee$^{1,\ddag}$}
\affiliation{$^{1}$Department of Physics, Massachusetts Institute
of Technology, Cambridge, Massachusetts 02139, USA}\affiliation{$^{2}$Center for Materials Science and Engineering, Massachusetts Institute of Technology, Cambridge, Massachusetts 02139, USA}
\date{\today}

\begin{abstract}
We report thermodynamic measurements of the S=$\frac{1}{2}$ kagome lattice antiferromagnet ZnCu$_{3}$(OH)$_{6}$Cl$_{2}$, a promising candidate system with a spin-liquid ground state. Using single crystal samples, the magnetic susceptibility both perpendicular and parallel to the kagome plane has been measured.  A small, temperature-dependent anisotropy has been observed, where $\chi_{z}/ \chi_{p} > 1$ at high temperatures and $\chi_{z}/ \chi_{p} < 1$ at low temperatures.  Fits of the high-temperature data to a Curie-Weiss model also reveal an anisotropy.  By comparing with theoretical calculations, the presence of a small easy-axis exchange anisotropy can be deduced as the primary perturbation to the dominant Heisenberg nearest neighbor interaction.  These results have great bearing on the interpretation of theoretical calculations based on the kagome Heisenberg antiferromagnet model to the experiments on ZnCu$_{3}$(OH)$_{6}$Cl$_{2}$.
\end{abstract}

 \maketitle

\begin{figure}
\centering
\includegraphics[width=8.7cm]{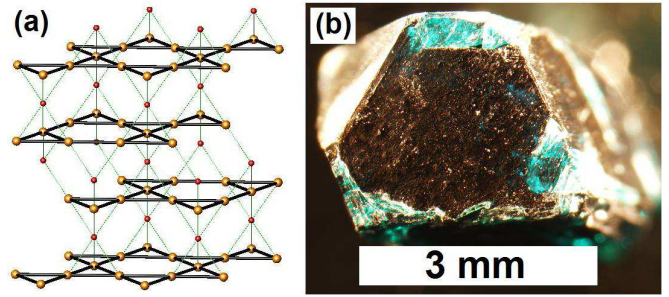} \vspace{-4mm}
\caption{(color online) (a) Structure of herbertsmithite with Cu$^{2+}$(big brown spheres) and Zn$^{2+}$(small red spheres) displayed. The Cu-Cu bonds (thick black solid lines) are all equivalent as are the Cu-Zn bonds (thin green dotted lines). (b) The oriented single crystal sample (mass = 55.5~mg) of herbertsmithite used in the magnetic susceptibility measurements.} \vspace{-4mm}
\label{Figure1}
\end{figure}

The quantum spin liquid, a fundamentally new state of matter whose ground state does not break conventional symmetries, has generated much interest in condensed matter physics\cite{Anderson1973,Anderson1987}. It has long been realized that the S=$\frac{1}{2}$ Heisenberg antiferromagnet on the kagome lattice (composed of corner sharing triangles) is an ideal system in which to look for spin-liquid physics due to the high degree of frustration, small spin, and low dimensionality. Herbertsmithite, the $x=1$ end member of the family Zn paratacamite [Zn$_{x}$Cu$_{4-x}$(OH)$_{6}$Cl$_{2}$], is arguably one of the best candidate systems to study quantum spin liquids.\cite{Shores} With weak interplane coupling, it consists of kagome planes of Cu$^{2+}$ ions separated by layers of nonmagnetic Zn$^{2+}$ ions, depicted in Fig.~\ref{Figure1}(a). The current experimental evidence is consistent with the presence of a spin-liquid ground state in this material\cite{Helton,Mendels,Wulferding,deVries}. The Hamiltonian of herbertsmithite consists of a Heisenberg exchange term, with possible perturbations such as a Dzyaloshinskii-Moriya (DM) interaction\cite{Grohol,Elhajal,Rigol1,Rigol2} and exchange anisotropy\cite{Starykh}. With a Cu-O-Cu antiferromagnetic superexhange interaction of approximately 17~meV, no magnetic transition or long range ordering has been observed down to $T=50$~mK\cite{Helton,Mendels,Imai}. It is important to perform measurements on single crystal samples so that comparisons can be made to theoretical calculations assuming different perturbations to the Hamiltonian\cite{Rigol1,Rigol2}, such as a DM interaction and exchange anisotropy, to determine the presence and magnitude of such perturbations.  Resolving this issue is all the more pressing in light of recent theoretical work which strongly points to a spin-liquid ground state for the S=1/2 kagome lattice with isotropic (Heisenberg) exchange.\cite{White}

Recently, large single crystal samples of the paratacamite family, including herbertsmithite, have been synthesized\cite{Han}. A powder sample was first synthesized inside a sealed quartz tubing and transported under a temperature gradient in a three zone furnace for crystallization. The high quality of the crystals was confirmed by inductively coupled plasma metal analysis, x-ray diffraction, neutron diffraction, polarized optics and thermodynamic measurements. Anomalous synchrotron x-ray diffraction confirmed the absence of antisite disorder where Zn$^{2+}$ ions appear on the Cu sites in the kagome layer\cite{Freedman}.  Rather, the main source of disorder is the presence of a small fraction of excess Cu$^{2+}$ ions within the Zn interlayers.   Raman spectroscopy provides further support of a gapless spin-liquid ground state\cite{Wulferding} while $\mu$SR points to an easy-axis anisotropy parallel to the c axis for magnetization\cite{Ofer3}. In this Letter, the magnetic susceptibility and specific heat have been investigated with fields applied both within and normal to the kagome plane. The roles of an easy-axis exchange anisotropy, a Dzyaloshinskii-Moriya interaction, and an anisotropic $g$ factor for the Cu magnetic moment are discussed.

\begin{figure}
\centering
\includegraphics[width=7.8cm]{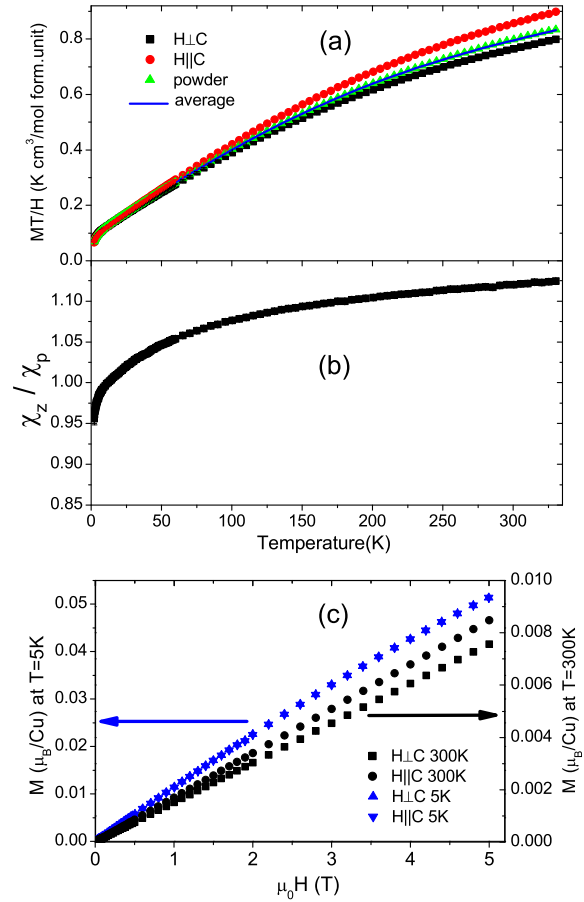} \vspace{-4mm}
\caption{(color online) (a) Magnetic susceptibility, plotted as $MT/H = \chi T$, measured under an applied field of $\mu_{0}H$~=~1~T which was oriented both perpendicular to and along the c-axis.  The data from a powder sample are also plotted and compared to the ``average'' of the single crystal results, as described in the text. (b) The anisotropy $\chi_{z}$/$\chi_{p}$ of the measured susceptibility plotted as a function of temperature. (c) Magnetization versus field measurements at $T=5$~K and 300~K. The vertical scale for each temperature is indicated by the arrow.} \vspace{-4mm}
\label{Figure2}
\end{figure}

Magnetic susceptibility measurements were performed on a 55.5~mg single crystal sample of herbertsmithite (2.3 mm $\times$ 2.5 mm $\times$ 2.7 mm), shown in Fig.~\ref{Figure1}(b), using a SQUID magnetometer (Quantum Design).  The nearly cubic shape of the sample minimizes a demagnetization correction to the measurements, allowing for a clean measurement of the intrinsic anisotropy of the material.  The crystalline axes and the narrow mosaic of the sample were confirmed using an x-ray diffractometer equipped with an area detector.  A plastic holder was designed and made for securing the crystal for susceptibility measurements with magnetic field applied perpendicular ($\chi_{p}$) or parallel ($\chi_{z}$) to the crystalline $c$ axis. The background from the plastic holder was measured to be negligibly small relative to the signal from the sample. In Fig.~\ref{Figure2}(a), the quantities $\chi_{z}T$ and $\chi_{p}T$ for temperatures between 2 and 330~K are plotted (where we assume $\chi = M/H$ in the linear regime). The quantity $\chi_{powder}T$, measured on a polycrystalline collection of several dozen random orientated crystals from the same batch, is plotted along with $\chi_{average}T=\frac{1}{3}$($\chi_{z}+2\chi_{p})T$, the calculated powder average. The latter two collapse onto the same curve as expected, pointing to the reliability of the single crystal measurements.

In Fig.~\ref{Figure2}(b), the anisotropy ratio of the magnetic susceptibility calculated as $\chi_{z}$/$\chi_{p}$ is plotted. As temperature is increased from 2 to 330~K, the ratio increases from 0.96 to 1.12 monotonically. The presence of anisotropy in the susceptibility agrees qualitatively with susceptibility measurements on aligned powders\cite{Ofer2} and recent $\mu$SR measurements on single crystals\cite{Ofer3}. In Fig.~\ref{Figure2}(c), magnetization measurements taken at $T=5$ and 300~K are plotted as a function of applied field. At $T=5$K, the anisotropy ratio is close to unity and the two curves overlap for the entire field range. At $T=300$~K, there is a clear anisotropy with the $c$ axis being the higher susceptibility direction. The observed magnetic anisotropy is independent of the applied field.

\begin{figure}
\centering
\includegraphics[width=7.8cm]{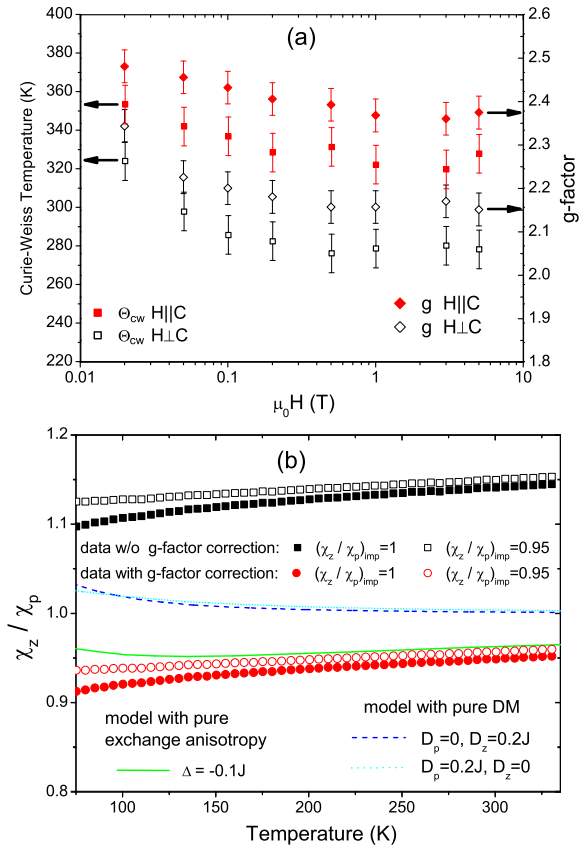} \vspace{-4mm}
\caption{(color online) (a) Curie-Weiss temperatures and $g$ factors calculated from fits of the magnetic susceptibilities between $T=150$ and 330~K, as described in the text. The proper vertical scale for each data set is indicated by the arrow. (b) The susceptibility anisotropy ratio of the intrinsic kagome spins after subtracting the impurity contribution, with and without a correction for the $g$-factor anisotropy, as described in the text. In the model for subtracting the impurity contribution, temperature independent anisotropy ratios $(\chi_{z}/\chi_{p})_{imp}$ were assumed.   The three curves represent exact diagonalization calculations\cite{Rigol2} for the anisotropy ratio considering the effects of an easy-axis exchange anisotropy and a DM interaction separately.}  \vspace{-4mm}
\label{Figure3}
\end{figure}

The high quality of the susceptibility data allows for further analysis to better understand the intrinsic behavior of the interacting spins on the kagome layers.  The primary results of this paper are shown in Fig.~3.  For the susceptibility data at high temperatures, Curie-Weiss fits were performed for each data set taken at various fields. The Curie-Weiss temperatures and $g$ factors determined from the fits (which take into account the corrections based on high-temperature series expansion calculations\cite{Harris,Matan}) are plotted in Fig.~\ref{Figure3}(a). For both field orientations, the Curie-Weiss temperatures and $g$ factors increase slightly upon lowering the applied field below $\mu_{0}H~\simeq~0.2$~T. At $\mu_{0}H~=~1$~T, the anisotropy ratio for the $g$-factor is $g_{z}/g_{p}=1.10$. A similar $g$ factor anisotropy, though slightly smaller, has been deduced from ESR work\cite{Zorko} on powders.

It is important to separate out the anisotropy of the Cu moments intrinsic to the kagome planes from that related to the impurity spins. The experimentally measured magnetic susceptibility originates from both the kagome plane and the weakly interacting Cu$^{2+}$ impurities on the interlayer sites. Assuming that the intrinsic kagome susceptibility becomes much smaller than the impurity contribution as $T \rightarrow 0$~K, consistent with recent NMR measurement on single crystal samples\cite{Imai2}, we model the impurity susceptibility with a Curie-Weiss law where $\Theta_{CW} \simeq 1.3$~K\cite{Bert}.  The best fit gives an estimated 10\% Cu$^{2+}$ ions which occupy the interlayer sites for this nondeuterated sample.  Then, by assuming a temperature independent anisotropy for the impurities, the impurity contribution to the susceptibility can be subtracted revealing the anisotropy of the intrinsic kagome spins.  The only remaining free parameter is the anisotropy ratio for the impurities $(\chi_{z}$/$\chi_{p})_{imp}$, and in our analysis, we use the value $(\chi_{z}$/$\chi_{p})_{imp}=1$.

The deduced anisotropy of the susceptibility for the intrinsic kagome spins is plotted in Fig.~\ref{Figure3}(b).  The main observation, which is relatively insensitive to the model parameters, is that $\chi_{z}$/$\chi_{p}$ for the intrinsic susceptibility is a monotonically increasing function of temperature for $T>150$~K.  This provides useful information on the importance of additional terms in the spin Hamiltonian, as we discuss further below.  Moreover, since we have deduced the anisotropy of the $g$ factor resulting from the Curie-Weiss analysis, we can correct for this in determining the $\chi_{z}$/$\chi_{p}$ ratio for the intrinsic kagome spins.  The $g$-factor corrected data are also plotted in Fig.~3(b).  At low temperatures (below $\sim 5$~K) where the impurity contribution dominates the susceptibility, the measured ratio for $\chi_{z}$/$\chi_{p}$ is actually less than 1.  If we assume a value of $(\chi_{z}$/$\chi_{p})_{imp}=0.95$, the deduced anisotropy ratio for the intrinsic kagome spins exhibits a slight upturn as the temperature is cooled below $T\approx100$~K.  However, the main conclusions of our analysis based on the data for temperature above $T=150$~K are not quantitatively changed.

\begin{figure}
\centering
\includegraphics[width=6.6cm]{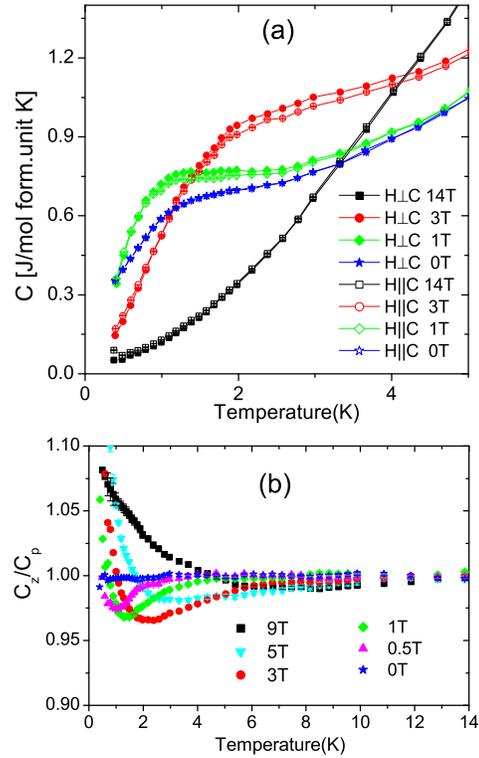} \vspace{-4mm}
\caption{(color online) (a) Low temperature specific heat data on a single crystal sample of herbertsmithite measured under various applied fields with two orientations. (b) The anisotropy ratio of the specific heat measured in the two field orientations.} \vspace{-4mm}
\label{Figure4}
\end{figure}

Our experimental results shed light on the roles played by various perturbations to the spin Hamiltonian beyond the nearest neighbor Heisenberg model. The observed anisotropy of the intrinsic susceptibility can be compared with theoretical calculations using 15-site exact diagonalization (ED) by Rigol and co-workers\cite{Rigol2}. If an easy-axis exchange anisotropy $H_{EA}$=$\Delta$$\Sigma$$_{i,j}$($S_{i}^{x}S_{j}^{x}$ + $S_{i}^{y}S_{j}^{y}$) with $\Delta<$ 0 is considered, the shape of the anisotropy versus temperature curve matches our measurements over a wide temperature range, as shown in Fig.3(b). In fact, comparing our $g$-factor corrected data with the calculation with $\Delta = -0.1$~J gives a good match for the slope for $T>150$~K as well as the magnitude for $\chi_{z}/\chi_{p}$. The presence of an anisotropic exchange is consistent with recent $\mu$SR measurements on single crystal herbertsmithite\cite{Ofer3} and work on partially aligned powders
\cite{Ofer2}.  The difference in the Curie-Weiss temperatures for the two field orientations is also consistent with the deduced magnitude of the easy-axis exchange anisotropy.  That is, in Fig.3~(a), $\Theta_{CW}$ for the field along the $c$ axis is larger than that for the field within the kagome plane by about 10\%, as one would expect for an easy-axis exchange anisotropy of $\Delta \approx -0.1$~J.

The DM interaction, $H_{DM}$=$\Sigma$$_{i,j}D_{z}$($\vec{S}_{i}$ $\times$ $\vec{S}_{j}$)$_{z}$ + $\vec{D}_{p}$ $\cdot$ ($\vec{S}_{i}$ $\times$ $\vec{S}_{j}$), has a much smaller effect on the anisotropy ratio\cite{Rigol2}. For a wide range of $D_{z}$ and $D_{p}$ values (the out-of-plane and the in-plane components of the DM vector, respectively), the primary effect is to slightly increase the anisotropy ratio from unity, where $\chi_{z}$/$\chi_{p}$ monotonically decreases as temperature increases.  The results of two model calculations which only include a DM term are plotted (one with $D_z = 0.2 J$ and one with $D_p = 0.2 J$). Our data appear to rule out such scenarios where only a DM term is present, as a small easy-axis exchange anisotropy is needed to give the observed $\chi_{z}$/$\chi_{p} < 1$ as well as the temperature dependence.

The specific heat was measured on a 4.10~mg single crystal sample of herbertsmithite using a Quantum Design physical property measurement system (PPMS). The sample was prepared so that its orientation could be changed $in$ $situ$ without remeasuring the background. The specific heat was measured in two field orientations: with the field oriented in the kagome plane $C_{p}$ and perpendicular to the plane $C_{z}$.  The data for a wide range of applied fields up to $\mu_{0}H$~=~14~T are plotted in Fig.~\ref{Figure4}(a). The ratio $C_{z}/C_{p}$ is plotted in Fig.~\ref{Figure4}(b) which reveals a small magnetocaloric anisotropy. As a check of systematic errors in the  measurement, the data measured in zero field, taken with both crystal orientations, collapse onto the same line. At temperatures higher than 15~K (not shown), the specific heat under all applied magnetic fields is identical for both orientations, within error.

At the lowest temperatures (below $\sim 10$~K), it is likely that the magnetocaloric anisotropy derives from the impurities. The observed $C_{z}/C_{p}>$ 1 for $k_B T \ll \mu_0 H$ indicates that the impurities are easier to polarize with an in-plane field. This idea is further supported by the observation in Fig.~2(b) that the ratio $\chi_{z}/\chi_{p}$ begins to decrease very rapidly upon cooling below $T=5$~K.  This indicates the impurity moments develop a $g$-factor anisotropy with $(\chi_{z}$/$\chi_{p})_{imp} < 1$ at low temperatures.  As another possibility, it has been shown that the presence of a DM interaction can mix the triplet and singlet states\cite{Rigol1} so that the total spin is not a good quantum number. The observed field independence of the anisotropy for the susceptibility coupled with the field dependence of anisotropy of the specific heat point to the possibility that the singlet states may be coupled to the applied field. Further theoretical calculations would be useful to determine how the thermodynamic quantities should behave under the application of in-plane and out-of-plane fields.

In summary, we have measured the anisotropy of the magnetic susceptibility in a single crystal sample of herbertsmithite.  The temperature dependence of the anisotropy allows one to deduce the important additional terms in the spin Hamiltonian beyond nearest neighbor Heisenberg exchange.  A comparison with previous exact diagonalization calculations indicates the presence of an easy-axis exchange anisotropy with $\Delta \approx -0.1$~J. This small value for the anisotropy indicates that the Heisenberg model is a reasonable approximation to define the physics of herbertsmithite.  However, calculations starting from the Ising limit and approaching the Heisenberg limit for the S=1/2 kagome antiferromagnet may provide useful insight into the behavior of herbertsmithite.  A field- and temperature-dependent anisotropy in the specific heat measured in different field orientations is also observed. Further theoretical calculations which include an anisotropic exchange interaction as well as a DM interaction would be most useful for a detailed comparison with the data.

We thank D. G. Nocera, A. Keren, J. S. Helton, M. Rigol, S. Todadri, and P. Lee for useful discussions.  This work was supported by the Department of Energy (DOE) under Grant No. DE-FG02-07ER46134.

$^{\ddag}$email: tianheng@alum.mit.edu, younglee@mit.edu
\bibliography{Anisotropy}
\end{document}